\documentclass[prd,onecolumn, nofootinbib, 10pt]{revtex4}  
\usepackage{graphicx}
\usepackage{amsmath}
\usepackage{amsfonts}
\usepackage{amssymb}
\usepackage{bm}
\usepackage{appendix}
\usepackage{mathtools}
\usepackage{comment}
\usepackage{bbold}
\usepackage{color}
\usepackage{slashed}
\usepackage{hyperref}
\usepackage{subfigure}
\usepackage{setspace}
\usepackage{enumitem}
\usepackage{longtable}
\usepackage{wasysym}
\usepackage[usenames,dvipsnames]{xcolor}
\usepackage{bm}
\usepackage{multirow}
\usepackage[letterpaper, margin=.8in, top=0.85in, bottom=0.85in]{geometry}

\pdfoutput=1

\begin{document}

\title{Model-Independent Extraction of $|V_{cb}|$ from $\bar{B}\rightarrow D^* \ell \overline{\nu}$}
\author{Benjam\'{i}n Grinstein}
\affiliation{Physics Department, University of California, San Diego, La Jolla, CA 92093, USA}
\author{Andrew Kobach}
\affiliation{Physics Department, University of California, San Diego, La Jolla, CA 92093, USA}

\date{\today}

\begin{abstract}
We fit the unfolded data of $\bar{B}^0\rightarrow D^{*+} \ell \overline{\nu}$ from the Belle experiment, where $\ell \equiv e, \mu$, using a method independent of heavy quark symmetry to extrapolate to zero-recoil and extract the value of $|V_{cb}|$.  This results in $|V_{cb}| = (41.9^{~+2.0}_{~-1.9})\times 10^{-3}$, which is robust to changes in the theoretical inputs and very consistent with the value extracted from inclusive semileptonic $B$ decays.  
\end{abstract}

\maketitle

\linespread{1}

\section{Introduction}

The discrepancy between the measured values of the CKM matrix element $|V_{cb}|$ from inclusive versus exclusive semileptonic $B$ decays  has been an ongoing dilemma for a few decades; for a recent review, see Ref.~\cite{Ricciardi:2016pmh}.  Currently, the world averages for $|V_{cb}|$ are~\cite{Amhis:2016xyh,Bauer:2004ve}:
\begin{eqnarray}
|V_{cb}| &=& (39.18 \pm 0.99) \times 10^{-3} \hspace{0.2in} (\bar{B}\rightarrow D\ell\overline{\nu}) \\
|V_{cb}| &=& (38.71 \pm 0.75) \times 10^{-3} \hspace{0.2in} (\bar{B}\rightarrow D^*\ell\overline{\nu}) \\
|V_{cb}| &=& (42.19 \pm 0.78) \times 10^{-3} \hspace{0.2in} (\bar{B}\rightarrow X_c\ell\overline{\nu}, \text{ kinetic scheme}) \\ 
|V_{cb}| &=& (41.98 \pm 0.45) \times 10^{-3} \hspace{0.2in} (\bar{B}\rightarrow X_c\ell\overline{\nu}, \text{ 1S scheme}) 
\end{eqnarray}
The extracted values of $|V_{cb}|$ from exclusive decays, especially the value of $|V_{cb}|$ measured in $\bar{B}\rightarrow D^*\ell\overline{\nu}$, are systematically lower than those from inclusive decays.  When measuring $|V_{cb}|$ in exclusive $\bar{B}\rightarrow D^{(*)}\ell\overline{\nu}$ decays, it has been commonplace to use the parameterization developed by Caprini, Lellouch, and Neubert (CLN)~\cite{Caprini:1995wq,Caprini:1997mu}, which utilizes not only dispersion relations to bound the hadronic form factors, but also relations at $1/m_Q$ in the heavy quark expansion, and claims to describe the full differential decay to within a few percent given only 4 parameters.  Interestingly, it has now become clear that the CLN parameterization is no longer a good fit to the $\bar{B}\rightarrow D\ell\overline{\nu}$ data from current experiments~\cite{Bigi:2016mdz}.  One would naively expect the same to be true, in general, for $\bar{B}\rightarrow D^*\ell\overline{\nu}$ as well.   

A recent analysis by Belle~\cite{Abdesselam:2017kjf} used the CLN parameterization to extract a
value for $|V_{cb}|$ at zero recoil given a dataset of $\bar{B}^0\rightarrow D^{*+} \ell
\overline{\nu}$, resulting in $|V_{cb}| = (37.4 \pm 1.3)\times10^{-3}$. This is one of the first
times an experiment with a large dataset has published unfolded data for $\bar{B}\rightarrow
D^*\ell\overline{\nu}$, so one can easily explore if a different parameterization of the hadronic
form factors gives rise to a different value of $|V_{cb}|$.  We use the method developed by Boyd,
Grinstein, and Lebed (BGL) in Refs.~\cite{Boyd:1994tt, Boyd:1995cf, Boyd:1995sq, Boyd:1995tg,Boyd:1997kz}, from
which the original CLN parameterization was based, but which does not utilize any assumptions from
heavy quark symmetry.\footnote{A brief history of the developments that led to the BGL parametrization and a more complete list of references can be found in Ref.~\cite{Grinstein:2015wqa}.}   
We find that while both the CLN and BGL parameterizations give a good fit to the Belle data, the BGL parameterization gives a larger value for $|V_{cb}|$ when extrapolating to zero recoil, consistent with the value of $|V_{cb}|$ measured in inclusive analyses~\cite{Amhis:2016xyh, Bauer:2004ve}. 
As we prepared this manuscript, a similar analysis was made public, using the same BGL parameterization~\cite{Bigi:2017njr}.  We are pleased to see the results match nearly identically.  Another analysis was recently made public, which highlighted the importance of including theoretical uncertainties from $\Lambda_\text{QCD}/m_Q$ corrections in the relations between the form factors in the heavy quark expansion~\cite{Bernlochner:2017jka}, which suggests that the uncertainties associated with the extracted value of $|V_{cb}|$ using the CLN method may be underestimated.   Moreover, in their analysis of $B\to D\ell\nu$ lattice data, the Fermilab Lattice and MILC collaborations decide not to quote results of their  CLN fits  because they  are more confident in the errors obtained from the  BGL parameterization which, they state,  can be used to obtain $|V_{cb}|$ even as the uncertainties become arbitrarily more precise~\cite{Lattice:2015rga}.

In all, the current evidence points to the strong possibility that the tension between inclusive and exclusive measurements of $|V_{cb}|$ in semileptonic $B$ decays may be due to the use of the CLN parameterization of $\bar{B}\rightarrow D^{(*)} \ell \overline{\nu}$, and a different parameterization should be employed.

\section{Differential Decay Rate and BGL Parameterization}
\label{sec:2}

Using the notation in Ref.~\cite{Boyd:1997kz}, the $\bar{B}\rightarrow D^*$ matrix elements are defined as
\begin{eqnarray}
\label{matel1}
\langle D^*(\varepsilon, p') | \bar{c} \gamma^\mu b | \bar{B}(p) \rangle &=& ig \epsilon^{\mu\nu\alpha\beta} \varepsilon^*_\nu p_\alpha p'_\beta , \\
\label{matel2}
\langle D^*(\varepsilon, p') | \bar{c} \gamma^\mu \gamma^5 b | \bar{B}(p) \rangle &=& f \varepsilon^{*\mu} + (\varepsilon^*\cdot p) [ a_+ (p+p')^\mu + a_- (p-p')^\mu ] ,
\end{eqnarray}
where $\varepsilon^\mu$ is the polarization tensor of the vector $D^*$ meson.  In the limit when the final-state leptons are massless, the full differential decay rate for $\bar{B}\rightarrow D^* \ell \overline{\nu}$ is
\begin{eqnarray}
\label{diffdecay}
\frac{d\Gamma(\overline{B}\rightarrow D^* \ell \overline{\nu})}{dw~d\cos\theta_\ell~d\cos\theta_v~d\chi} &=& \frac{3\eta_{ew}^2 G_F^2 |V_{cb}|^2 }{1024\pi^4} |{\bf p}_{D^*}| q^2 r \bigg( (1-\cos\theta_\ell)^2 \sin^2\theta_v H_+^2 + (1+\cos\theta_\ell)^2 \sin^2\theta_v H_-^2 \nonumber \\
&& \hspace{0.5in}+ ~4\sin^2\theta_\ell \cos^2\theta_v H_0^2 - 2\sin^2\theta_\ell \sin^2\theta_v \cos2\chi H_+ H_- \nonumber \\
&& \hspace{0.5in} - ~4\sin\theta_\ell(1-\cos\theta_\ell)\sin\theta_v \cos\theta_v \cos\chi H_+ H_0 \nonumber \\
&& \hspace{0.5in}+ ~4 \sin\theta_\ell(1+\cos\theta_\ell)\sin\theta_v \cos\theta_v \cos\chi H_- H_0 \bigg) ,
\end{eqnarray}
where $q^\mu$ is the 4-momentum of the lepton system, $r\equiv m_{D^*}/m_B$, and $|{\bf p}_{D^*}|$ is the magnitude of the $D^*$ 3-momentum in the rest frame of the $\bar{B}$:
\begin{equation}
w \equiv \frac{m_B^2 + m_{D^*}^2 - q^2}{2m_B m_{D^*}} , \hspace{0.5in}q^2  = m_B^2 + m_{D^*}^2 - 2m_B m_{D^*} w, \hspace{0.5in} |{\bf p}_{D^*}| = m_{D^*} \sqrt{w^2-1} .
\end{equation}
Here, $H_+$, $H_-$, and $H_0$ are form factors associated with each of the three helicity states of the $D^*$, all of which are functions of $q^2$.  Also, $\theta_\ell$ is the angle between the anti-neutrino and the direction antiparallel to the $D^*$ in the rest frame of the leptonic system, $\theta_v$ is the angle between the $D^*$ momentum and its daughter $D$ meson, and $\chi$ is the angle between the planes defined by the the leptonic system and the $D^*$ system.  The factor $\eta_{ew}$ incorporates the leading electroweak corrections~\cite{Sirlin:1981ie}, $\eta_{ew} = 1 + \alpha/\pi \ln (M_Z/m_B) \simeq 1.0066$.  
In terms of the form factors in Eqs.~(\ref{matel1}) and~(\ref{matel2}),
\begin{eqnarray}
\label{formfactrans}
H_+ &=& f - m_B |{\bf p}_{D^*}| g , \\
H_- &=& f + m_B |{\bf p}_{D^*}| g , \\
\label{H0formfactor}
H_0 &=& \frac{1}{m_{D^*}\sqrt{q^2}} \bigg[ 2m_B^2 |{\bf p}_{D^*}|^2 a_+ - \frac{1}{2} \big( q^2 -m_B^2 + m_{D^*}^2\big) f \bigg] \equiv \frac{\mathcal{F}_1}{\sqrt{q^2}} .
\end{eqnarray}

A detailed discussion about the BGL method for parameterizing the form factors $f$, $g$, and
$\mathcal{F}_1$ can be found in Ref.~\cite{Boyd:1997kz}. The final result gives a parametrization of
each form factor in terms of $N+1$ coefficients: 
\begin{eqnarray}
\label{formfactors}
g(z) &=& \frac{1}{P_g(z) \phi_g(z)} \displaystyle\sum_{n=0}^N a_n z^n , \hspace{0.25in} f(z) = \frac{1}{P_f(z) \phi_f(z)} \displaystyle\sum_{n=0}^N b_n z^n , \hspace{0.25in} \mathcal{F}_1(z) = \frac{1}{P_{\mathcal{F}_1}(z) \phi_{\mathcal{F}_1}(z)} \displaystyle\sum_{n=0}^N c_n z^n,
\end{eqnarray}
where the conformal variable $z$ is defined as
\begin{equation}
z \equiv \frac{\sqrt{w+1}-\sqrt{2a}}{\sqrt{w+1}+\sqrt{2a}}\, .
\end{equation}
Here, $a=1$ can be chosen such that $z=0$ corresponds to zero recoil, and the coefficients $a_n$,
$b_n$, and $c_n$ are bounded by unitarity \cite{Boyd:1995sq},
\begin{equation}
\label{unitarybounds}
\sum_{n=0}^N |a_n|^2\le1,\qquad  \text{and}\qquad  \sum_{n=0}^N\left(|b_n|^2+|c_n|^2\right)\le1,\qquad  
\end{equation}  
From  Eq.~\eqref{H0formfactor},  $\mathcal{F}_1(0)=(m_B-m_{D^*})f(0)$; hence $b_0$ and $c_0$ are not
independent, i.e., 
\begin{equation} 
\label{c0const}
c_0 = \left(\frac{(m_B-m_{D^*})\phi_{\mathcal{F}_1}(0)}{\phi_f(0)} \right)b_0 .
\end{equation}
The Blaschke factors $P(z)$ remove poles for $  q^2< (m_B + m_{D^*})^2$  associated with on-shell production of $B_c^*$ bound states:
\begin{equation}
P_g(z) = \displaystyle\prod_i^4 \frac{z- z_{P_i}}{1-zz_{P_i}}, \hspace{0.25in} P_f(z) = P_{\mathcal{F}_1}(z) = \displaystyle\prod_i^4 \frac{z- z_{P_i}}{1-zz_{P_i}} ,
\end{equation}
\begin{equation}
z_P \equiv \frac{\sqrt{(m_B+m_{D^*})^2-m_P^2}-\sqrt{(m_B+m_{D^*})^2-(m_B-m_{D^*})^2}}{\sqrt{(m_B+m_{D^*})^2-m_P^2}+\sqrt{(m_B+m_{D^*})^2-(m_B-m_{D^*})^2}} .
\end{equation}
For the form factor $g$, the index $i$ runs over the 4 vector $B_c^*$ states, and for the $f$ and $\mathcal{F}_1$, it runs over the 4 axial vector states, as listed in Table~\ref{Bcmasses}.  The outer functions $\phi$ are defined as:
\begin{eqnarray}
\phi_g(z) &= &\sqrt{\frac{256n_I}{3\pi \chi^T(+u)}} \frac{r^2 (1+z)^2 (1-z)^{-1/2}}{[(1+r)(1-z)+2\sqrt{r}(1+z)]^4},  \\
\phi_f(z) &=& \frac{1}{m_B^2} \sqrt{\frac{16n_I}{3\pi \chi^T(-u)}}\frac{r (1+z) (1-z)^{3/2}}{[(1+r)(1-z)+2\sqrt{r}(1+z)]^4} , \\
\phi_{\mathcal{F}_1}(z) &=&  \frac{1}{m_B^3} \sqrt{\frac{8n_I}{3\pi \chi^T(-u)}}  \frac{r (1+z) (1-z)^{5/2}}{[(1+r)(1-z)+2\sqrt{r}(1+z)]^5} ,
\end{eqnarray}
where $n_I$ is the effective number of light quarks, $u\equiv m_c/m_b$, and $\chi^T(\pm u)$  is
related to a perturbative calculation at $q^2=0$.  We use the result for $\chi^T(\pm u)$ in the pole mass scheme, and take numerical values to be those listed in Table~\ref{inputs}, as done in Ref.~\cite{Boyd:1997kz}, ignoring the small contributions from condensates:
\begin{eqnarray}
\chi^T(+0.33) = 5.28\times10^{-4} \text{ GeV}^{-2}, \hspace{0.25in} \chi^T(-0.33) = 3.07\times10^{-4} \text{ GeV}^{-2} 
\end{eqnarray}

\begin{table}[h!]
\begin{tabular}{ | c | c |  }
\hline
Type & Mass  [GeV]  \\ \hline \hline
vector & 6.337, 6.899, 7.012, 7.280 \\
axial vector & 6.730, 6.736, 7.135, 7.142 \\ \hline
\end{tabular}
\caption{$B_c^*$ masses used in this analysis from~\cite{Eichten:1994gt}.  Only states that have a mass less than $m_B + m_{D^*}$ are included. }
\label{Bcmasses}
\end{table}

\begin{table}[h!]
\begin{tabular}{ | c | c |  }
\hline
$m_B$ & $5.279$ GeV \\ \hline
$m_{D^*}$ & $2.010 $ GeV \\ \hline
$\alpha_s$ & 0.22 \\ \hline
$m_b$ & 4.9 GeV \\ \hline
$m_c/m_b$ & 0.33 \\ \hline
$n_I$ & 2.6 \\ \hline
\end{tabular}
\caption{Numerical inputs for this analysis.   The meson masses are taken from Ref.~\cite{Olive:2016xmw}.  See the text at the end of Section~\ref{sec:2} for a discussion regarding the choices of the other parameters. }
\label{inputs}
\end{table}

We do not incorporate any uncertainties associated with the numerical inputs to the BGL theory
parameterization.  Specifically, the uncertainties associated with $m_B$ and $m_{D^*}$ are negligible in
this analysis. Also, varying the inputs that only affect the overall normalization of the outer function, namely
$n_I$ and those used in the perturbative calculation of $\chi^T$, amounts to changing the right-hand
side of the bounds in Eq.~\eqref{unitarybounds} by a corresponding fractional amount.
We impose the bounds Eq.~\eqref{unitarybounds} as constraints, however, the resultant best-fit values are orders of
magnitude from saturating the unitarity bounds, and these constraints effectively play no role.  
Because of this, changing the values of, say, $\alpha_s$, $m_b$, etc.,  will change the best-fit values for the free parameters in the theory, but will not change the extracted value $|V_{cb}|$, to a very good approximation. 
The remaining  inputs are  $B_c^*$ pole masses. They  affect both the effectiveness of the
unitarity bound and the shape of form factors. For example, for $F\equiv g,f,\mathcal{F}_1$, the unitarity bound implies that the truncation error  for finite $N$ in
Eq.~\eqref{formfactors} is bounded over the physical region $0\le z\le z_{\rm max}\approx0.056$ by
\[
\Delta F(z) <\frac{1}{|P_F(z_{\rm max})\phi_F(z_{\rm max})|}\frac{z^{N+1}_{\rm  max}}{\sqrt{1-z_{\rm max}^2}}\,.
\]
Since $P_F(z)$ varies by less than 15\% when the input pole masses are individually varied by 1\%,  the
ability to control the expansion is unaffected by uncertainties in pole masses.   Also, the precise location
of the pole masses affects the shape of the form factors in the physical region, but the parameters
of the fit accomodate small changes
in pole masses easily. For example, consider the $N=0$ parametrization of the form factor $f$.  Varying the value of the lowest pole mass by 1\%,  the change in $f$ is compensated by a
multiplicative factor of $1.07-0.17z$ to better than 1\% over the whole physical region.  For these reasons, 
we expect that while the values of the fit parameters may depend somewhat sensitively on the inputs,
the extracted value of $|V_{cb}|$ does not. We illustrate this by using the same input values as Ref.~\cite{Boyd:1997kz}, as listed in Tables~\ref{Bcmasses} and~\ref{inputs}, which are about 20 years old, but find that our extracted value of $|V_{cb}|$ is nearly identical to that of a recent analysis by the authors of Ref.~\cite{Bigi:2017njr}, which uses the current state-of-the-art theoretical inputs.   Extracting the value of $|V_{cb}|$ using the BGL parameterization is robust to changes in the numerical inputs. We discuss this more in Section~\ref{sec:fitting}.

\section{Fitting Results and Extracting $|V_{cb}|$} 
\label{sec:fitting}

\begin{figure}[h!]
\centering
\includegraphics[width=1.\textwidth]{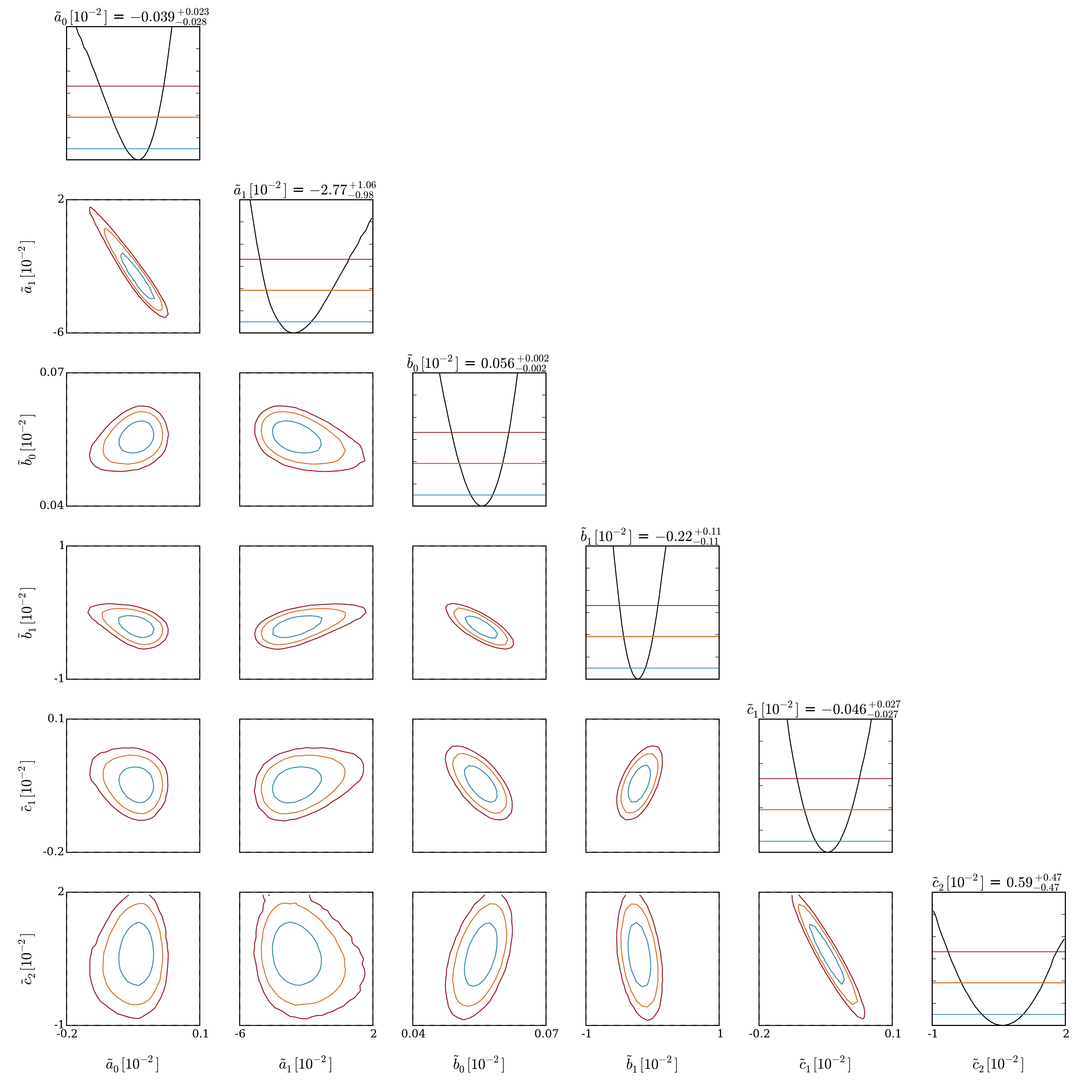}
\caption{The result of the fit to the form factors defined in Eq.~(\ref{formfactors}), where $\tilde{a}_i$, $\tilde{b}_i$, and $\tilde{c}_i$ are defined as $\eta_{ew} |V_{cb}| a_i$, $\eta_{ew} |V_{cb}| b_i$, and $\eta_{ew} |V_{cb}| c_i$, respectively. The blue, yellow, and red lines correspond to the $68.27\%$, $95\%$, and $99\%$ CL, respectively. Any jaggedness in the contours is from the finite size of the MCMC ensemble. }
\label{fitresults}
\end{figure}

\begin{figure}[h!]
\centering
\includegraphics[width=.85\textwidth]{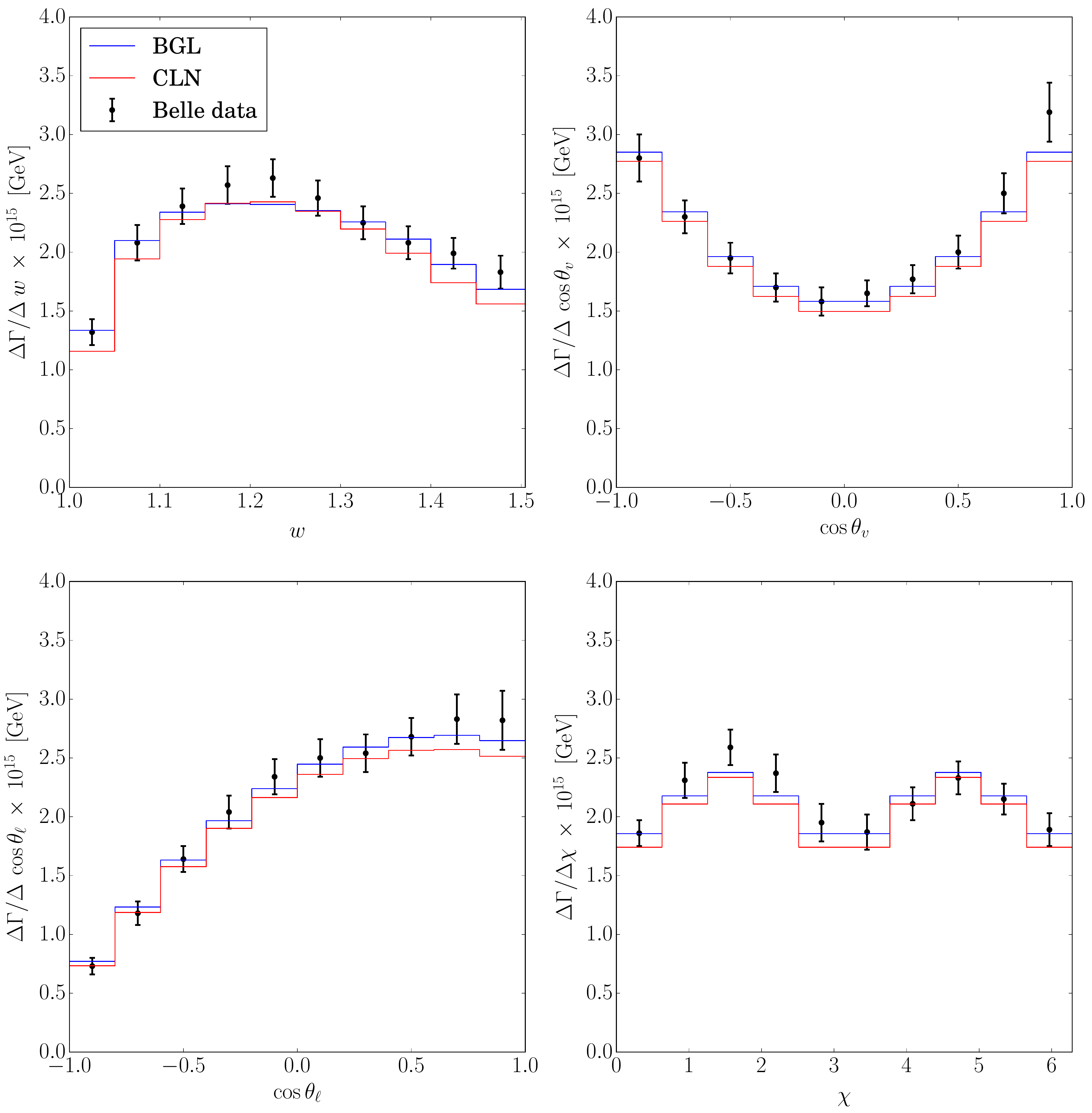}
\caption{ The Belle data (black points) compared to the results of our fit using the BGL parameterization (blue), and the results of the Belle analysis using the CLN parameterization (red). }
\label{belleplots}
\end{figure}

We fit to the unfolded data provided by Belle in Ref.~\cite{Abdesselam:2017kjf}, using the numerical inputs found in Tables~\ref{Bcmasses} and~\ref{inputs}.  We choose to truncate the series in Eq.~(\ref{formfactors}) at $\mathcal{O}(z^2)$ for $f$ and $g$, and at order $\mathcal{O}(z^3)$ for $\mathcal{F}_1$, as well as define the value $c_0$ in terms of $b_0$, as in Eq.~\eqref{c0const}.  This results in 6 free parameters, defined as $\tilde{a}_i \equiv  |V_{cb}| \eta_{ew} a_i$, $\tilde{b}_i \equiv|V_{cb}| \eta_{ew}  b_i$, where $i=0,1$, and $\tilde{c}_i\equiv  |V_{cb}|\eta_{ew}  c_i$, where $i=1,2$.  We choose to truncate to these orders because it results in a good fit to the data.  A Markov Chain Monte Carlo (MCMC) is used to estimate the shape of the likelihood and the shape around its extremum.  The result of the fit is shown in Fig.~\ref{fitresults}. We find $\chi^2_\text{min}/\text{dof} \simeq 27.7/34$.  Comparing to the Belle data, our best fit value for the BGL parameterization can be found in Fig.~\ref{belleplots}, compared to the fit performed by Belle using the CLN parameterization.  

\begin{figure}[h!]
\centering
\includegraphics[width=0.5\textwidth]{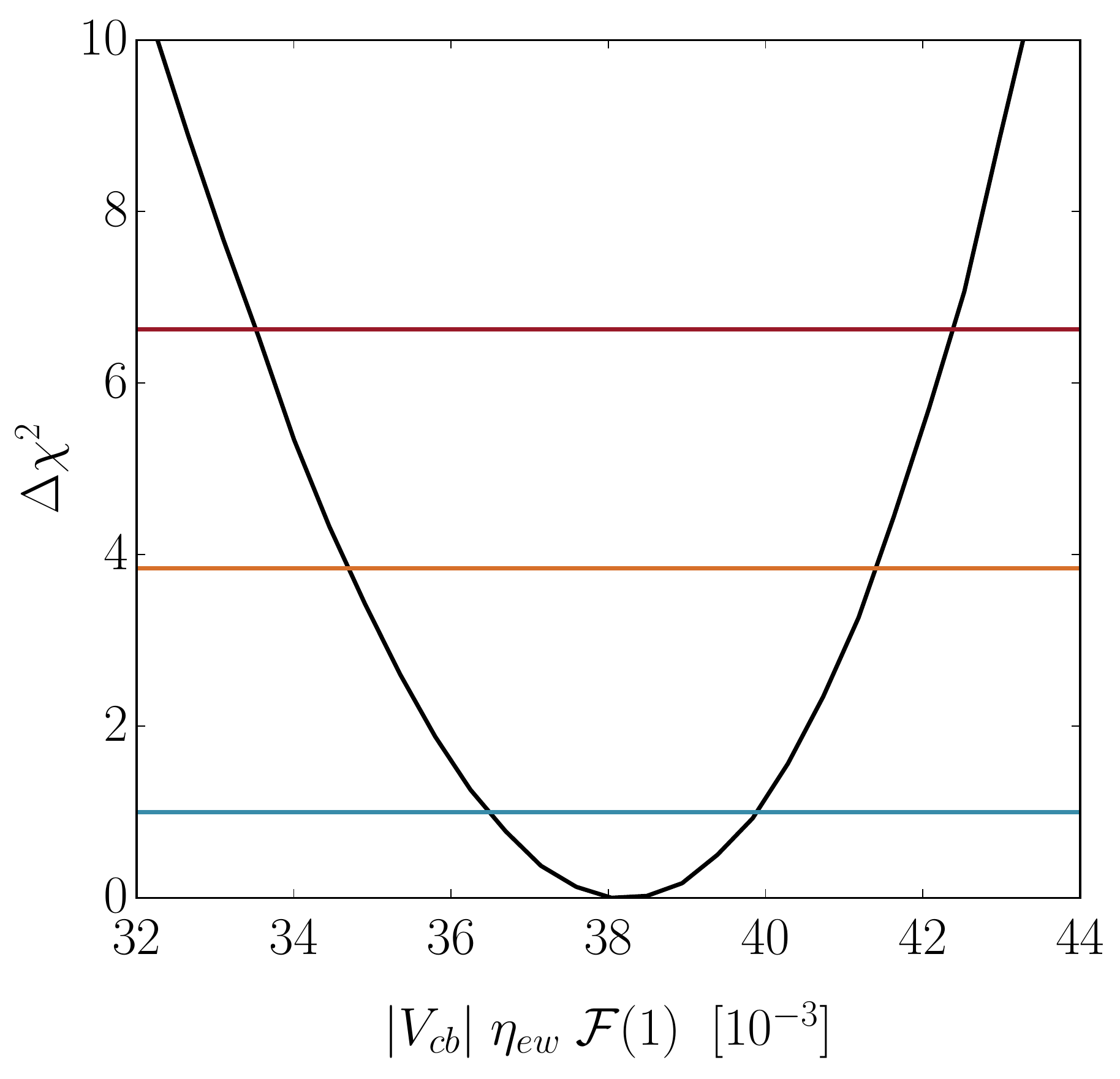}
\caption{The extracted value of $|V_{cb}| \eta_{ew} \mathcal{F}(1)$ from the Belle data. The blue, yellow, and red lines correspond to the $68.27\%$, $95\%$, and $99\%$ CL, respectively. }
\label{vcb}
\end{figure}

One requires input from the lattice to extract the value of $|V_{cb}|$.  The $\bar{B}\rightarrow D^* \ell \overline{\nu}$ rate is given
in terms of a function $\mathcal{F}(w)$ defined as \cite{Bailey:2014tva}:
\begin{eqnarray}
\frac{d\Gamma}{dw} &=& \frac{\eta_{ew}^2 G_F^2 |V_{cb}|^2 }{48\pi^3} (m_B - m_{D^*})^2 m_{D^*}^3 \sqrt{w^2-1} (w+1)^2  \nonumber \\
&& \hspace{0.5in} ~ \times \bigg[ 1+ \left(\frac{4w}{w+1} \right) \left(\frac{m_B^2+m_{D^*}^2-2wm_Bm_{D^*}}{(m_B-m_{D^*})^2} \right)  \bigg] |\mathcal{F}(w)|^2 .
\end{eqnarray}
One can estimate on the lattice the value of $\mathcal{F}(1)$.  Given the definition of the form factors in Eqs.~(\ref{diffdecay}), (\ref{formfactrans}), and (\ref{formfactors}), the relationship between $|V_{cb}| \eta_{ew} \mathcal{F}(1)$ and our fitting parameters is:
\begin{equation}
\label{master}
|V_{cb}| \eta_{ew} \mathcal{F}(1) = \frac{1}{2\sqrt{m_B m_{D^*}}} \left(\frac{|\tilde{b}_0|}{P_f(0) \phi_f(0)} \right)
\end{equation}
Given the statistical ensemble of $\tilde{b}_0$ provided by the MCMC, we can estimate the likelihood, and thus the variance in the $\chi^2$ about its minimum, i.e., $\Delta \chi^2$, for the value of $|V_{cb}|\eta_{ew}  \mathcal{F}(1)$ extracted from the Belle data, as shown in Fig.~\ref{vcb}.  This results in:
\begin{equation}
|V_{cb}| \eta_{ew}  \mathcal{F}(1) = (38.2^{~+1.7}_{~-1.6}) \times 10^{-3}
\end{equation} 
Using the  $\eta_{ew} = 1.0066$ and the FNAL/MILC value of $\mathcal{F}(1) = 0.906\pm0.013$ from Ref.~\cite{Bailey:2014tva}, the final result for $|V_{cb}|$ is 
\begin{equation}
|V_{cb}| = (41.9^{~+2.0}_{~-1.9}) \times 10^{-3}
\end{equation}
This result is nearly identical to that of the recent analysis by the authors of Ref.~\cite{Bigi:2017njr}, and systematically larger than the value extracted by the Belle experiment using the CLN parameterization~\cite{Abdesselam:2017kjf}.  Our result is comparable to a recent analysis by the authors of Ref.~\cite{Bernlochner:2017jka}, which reevaluate the CLN parameterization, using relations from heavy quark symmetry, but not ignoring important sources of theoretical uncertainty when fitting to the same data.  

As explained above, extracting the value of $|V_{cb}|$ using the BGL parameterization is robust to
changes in the numerical inputs.  We  explore further the effect of Blaschke factors on the
extraction of $|V_{cb}|$ by squaring the Blaschke factors for $f$ and $\mathcal{F}_1$. One might, a
priori,  expect to produce radical variations in the $z$ dependence of the form
factors.  Surprisingly, while we fit very different values for the free parameters, the extracted
value of $|V_{cb}|\eta_{ew}\mathcal{F}(1)$ is nearly unchanged (after updating the expression in
Eq.~(\ref{master}) to account for the different definition of the form factors). This gives further
evidence that the BGL parameterization is robust when extrapolating to zero recoil.  On the other hand, the precise values of the theoretical inputs may be essential when extracting the functional dependence of the form factors over the entire kinematic range.

\section{Summary and Conclusions}

The long-standing tension between the exclusive and inclusive determinations of $|V_{cb}|$ in semileptonic $B$ decays could have been exacerbated by underestimated uncertainties in the exclusive analyses when using the CLN parameterization~\cite{Bigi:2016mdz, Bernlochner:2017jka, Bigi:2017njr}.  As it stands, the data has become too precise to use the original CLN proposal in the case of $\bar{B}\rightarrow D \ell\overline{\nu}$, as discussed in Ref.~\cite{Bigi:2016mdz}, and one might expect the same could be true for $\bar{B}\rightarrow D^* \ell\overline{\nu}$.

The Belle experiment recently released an unfolded dataset of $\bar{B}^0\rightarrow D^{*+} \ell \overline{\nu}$, which resulted in the following value of $|V_{cb}|$ using the CLN parameterization~\cite{Abdesselam:2017kjf}:
\begin{equation}
|V_{cb}| = (37.4 \pm 1.3)\times10^{-3} \hspace{0.25in} \text{(CLN)}
\end{equation}
Using this  data, we fit to the BGL parameterization, which does not rely on any assumptions regarding heavy quark symmetry, and obtain:
\begin{equation}
|V_{cb}| = (41.9^{~+2.0}_{~-1.9}) \times 10^{-3} \hspace{0.25in} \text{(BGL)}
\end{equation}
This result is consistent with the value of $|V_{cb}|$ measured using inclusive semileptonic $B$ decays.  We are able to corroborate the results of the analysis in Ref.~\cite{Bigi:2017njr}, which appeared
as our manuscript was in preparation.  One can see by eye in Fig.~\ref{belleplots} that the BGL
  parametrization gives a different fit to the data, and a systematically larger value of the
  $\bar{B}\rightarrow D^* \ell \overline{\nu}$ rate near zero-recoil, than when using the CLN parameterization.  Furthermore, we find that the BGL method is robust when extrapolating to zero-recoil, i.e., large changes in the inputs associated with the form factors in Eq.~(\ref{formfactors}) result in almost no change in the extracted value of $|V_{cb}|$.

\begin{acknowledgements}
We thank Aneesh Manohar for useful conversations and feedback, Florian Bernlochner for providing the unfolded data from Belle, and Kevin Kelly for advice on graphics and using Monte Carlo.  This work is supported in part by DOE grant \#DE-SC0009919.
\end{acknowledgements}

\bibliography{bib}{}

\end{document}